\documentclass[aps,prl,preprint]{revtex4}
\usepackage{amsmath}
\usepackage{amsfonts}
\usepackage{amssymb}
\usepackage{graphicx}
\usepackage{color}	%added by RW for highlighting changes
\newcommand{\ket}[1]{\ensuremath{\left| #1 \right\rangle}}

\newcommand{\figref}[1]{Fig. \ref{#1}}

\begin{document}

\title{Is a single photon's wave front observable?}

\author{A.M. Zagoskin} 
\email{a.zagoskin@lboro.ac.uk}
\author{R.D. Wilson}\author{M. Everitt}\author{S. Savel'ev}\author{D.R. Gulevich}\author{J. Allen}
\affiliation{Department of Physics, Loughborough University, Loughborough, Leicestershire, LE11 3TU, United Kingdom}
\author{V.K. Dubrovich}
\affiliation{Special Astrophysical Observatory, Russian Academy of Sciences, and Cryogenic Nanoelectronics Laboratory, Nizhny Novgorod State Technical University, Russian Federation}
\author{E. Il'ichev}
\affiliation{Institute for Photonic Technologies, Jena, Germany}

\begin{abstract}
We propose a method of detecting the wave front of a single photon in several points more than a photon wavelength apart. It is based on the entangling interaction of the incoming photon with the quantum metamaterial sensor array, which produces the spatially correlated quantum state of the latter, and the quantum nondemolition readout of a collective observable (e.g., total magnetic moment), which characterizes this quantum state. We show that the effects of local noise (e.g., fluctuations affecting the elements of the array) are  suppressed relative to the signal from the spatially coherent field of the incoming photon. The realization of this approach in the microwave range would be especially useful and is within the reach of current experimental techniques.\end{abstract}

\maketitle

The ultimate goal and the theoretical limit of weak signal detection is the ability to detect a single photon against a noisy background. In this situation the inescapable noise produced by the measuring device itself may be the main threat, but the  uncertainty principle strongly restricts possible experimental techniques of increasing the signal-to-noise ratio. For example, a weak classical signal from a remote source can be distinguished from the local noise at the same frequency through its spatial correlations (using phase sensitive detectors; coincidence counters; etc) - i.e., by sensing its wave front. This method seems impossible in case of a single incoming photon, since it can only be absorbed one single time. Nevertheless such a conclusion would be too hasty. In this paper we show, that a combination of a quantum metamaterial\cite{Rakhmanov2008} (QMM)-based sensor array and quantum non-demolition\cite{Braginsky,Lupascu2007} (QND) readout of its quantum state allows, in principle, to detect a single photon in several points, i.e., to observe its wave front. 
Actually, there are a few possible ways of doing this, with at least one within the reach of current experimental techniques for the microwave range. The ability to resolve the quantum-limited signal from a remote source against a much stronger local noise would bring significant advantages to such diverse fields of activity as, e.g., microwave astronomy and missile defence.

Here the QMM array is modelled by a set of N qubits, which are coupled to two LC circuits: the one (A) represents the input mode, and the other (B) the readout. This model is closest to the case of microwave signal detection using superconducting qubits, which is both most feasible and most interesting (at least from the point of view of radioastronomy). Nevertheless our approach and conclusions apply generally, \textit{mutatis mutandis} (e.g., to the case of photonic crystal decorated with two level systems \cite{Greentree2008} or an array of SQUIDs \cite{Stiffell2005}). We will begin by discussing how such a detector system could work in principle. Before going on to demonstrate that a clear distinction between a single incident photon and the vacuum can be seen in the response of a simple two-qubit detector array using a fully quantum mechanical model. Then finally we will explore the role of inter-qubit coupling and increasing the size of the QMM array using a semi-classical mean field approach.

\begin{figure}[b!]
	\begin{center}
	\includegraphics[width=0.5\textwidth]{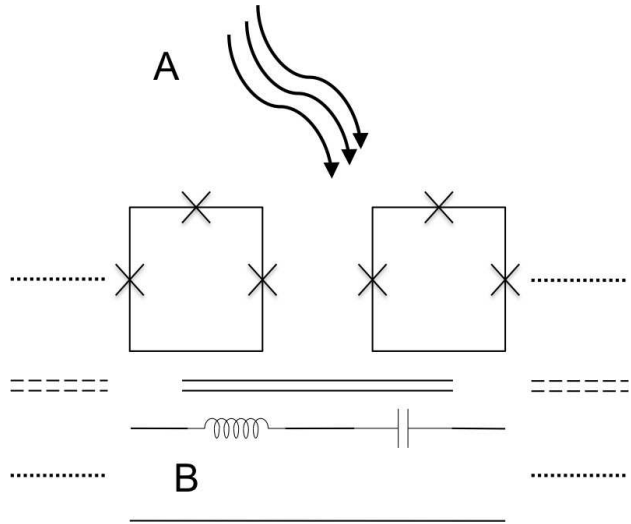} 
	\end{center}
\caption{Schematic for the photon detector system. Photons are incident on to the QMM array, which is comprised of $N$ qubits in this case. The QMM array is also coupled to the readout tank circuit in order to perform quantum non-deomolition measurement.}
\label{fig:schematic}
\end{figure}

The  system of \figref{fig:schematic} is described by the Hamiltonian
\begin{eqnarray}
H = H_a + V_a + H_{qb} + V_b + H_b.
\label{eq:1}
\end{eqnarray}
Here 
\begin{equation}
H_a = \omega_a (a^{\dag}a + 1/2) + f(t)(a^{\dag}+a)
\label{eq:1a}
\end{equation}
describes the input circuit, excited by the incoming field; 
\begin{equation}
H_{qb} = \left(-\frac{1}{2}\right)\sum_{j=1}^N \left(\Delta_j \sigma^x_j + \varepsilon_j \sigma^z_j\right)  
\label{eq:1qb}
\end{equation} 
is the Hamiltonian of the qubits; 
\begin{equation}
H_b = \omega_b (b^{\dag}b + 1/2) + h(t)(b^{\dag}+b)
\label{eq:1b}
\end{equation}
is the Hamiltonian of the output circuit with the probing field, used in case of IMT readout;
\begin{equation}
V_a = \sum_j g_j^a (a^{\dag}+a)\sigma^x_j, \:\: V_b = \sum_j g_j^b (b^{\dag}+b)\sigma^x_j
\label{eq:1V}
\end{equation}
describe the coupling between the QMM array and the input and output circuits. The effects of ambient noise can be taken into account, e.g., by adding an appropriate term $H_{\rm noise}$ to (\ref{eq:1}) or by including Lindblad operators in the master equation for the density matrix of the system (see, e.g., \cite{Zagoskin2011}).

%**********************************************************************************************************************************
%Numerical results by RW and DG from here

\textit{--Numerical results for two sensor qubits.} 	Consider the example of a QMM matrix comprised of two superconducting flux qubits   coupled to the input field and  a readout tank circuit - a setup close to the one already realized experimentally \cite{Shevchenko2008}. We take $\Delta=0$ and therefore the frequency of the flux qubits is simply given by $\epsilon$. In experiments this frequency is typically of the order of 10GHz and can be controlled by an applied external bias \cite{Mooij1999,Astafiev2010}.
	
	To ensure that the readout circuit feels the maximum effect of an incoming photon we choose its frequency to be on (or very near to) resonance with the input field, $\omega_{a}\approx\omega_{b}$. We also choose to work in the dispersive regime, where $\left|\omega_{a}-\epsilon\right|\gg g_{j}^{a}$ and $\left|\omega_{b}-\epsilon\right|\gg g_{j}^{b}$. This ensures that input field produces a Lamb shift in the detector qubits proportional to the number of incoming photons and prevents them from simply being absorbed by one of the qubits thus collapsing the photon wave function. This shift in the dynamics of the qubit array can then be detected by the readout circuit. Control of the frequencies of the qubits $\epsilon$ and the readout circuit $\omega_b$ gives this scheme a good degree of tunability. We choose coupling parameters $g_{j}^{a}/\epsilon=g_{j}^{b}/\epsilon=0.01$ in line with typical experiments \cite{Fink2008,Abdumalikov2008}. 

	In order to fully account for the effects of decoherence and measurement, we make use of the quantum state diffusion formalism \cite{Percival1998} to describe the evolution of the state vector \ket{\psi};
\begin{multline}
	\ket{d\psi}=-iH\ket{\psi}\mathrm{d}t+
	\sum_{j}\left[\left\langle\hat{L}_{j}^{\dag}\right\rangle\hat{L}_{j}-\frac{1}{2}\hat{L}_{j}^{\dag}\hat{L}_{j}
	-\frac{1}{2}\left\langle\hat{L}_{j}^{\dag}\right\rangle\left\langle\hat{L}_{j}\right\rangle\right]
	\ket{\psi}\mathrm{d}t+ \\
	\sum_{j}\left[\hat{L}_{j}-\left\langle\hat{L}_{j}\right\rangle\right]\ket{\psi}\mathrm{d}\xi_{j} ,
	\label{eqn:QSD}
\end{multline}
where $\ket{d\psi}$ and $\mathrm{d}t$ are the state vector and time increments respectively, $\hat{L}_{j}$ are the Lindblad operators and $\mathrm{d}\xi_{j}$ are the stochastic Wiener increments which satisfy $\overline{\mathrm{d}\xi_{j}^{2}}=\overline{\mathrm{d}\xi_{j}}=0$ and $\overline{\mathrm{d}\xi_{j}\mathrm{d}\xi_{j}^{*}}=\mathrm{d}t$. The effects of decoherence on the qubits are described by the Lindblad operators $L_{z}=\sqrt{2\Gamma_{z}}\sigma_{-}^{\left(i\right)}$ and $L_{xy}=\sqrt{2\Gamma_{xy}}\sigma_{+}^{\left(i\right)}\sigma_{-}^{\left(i\right)}$ acting on both qubits. These operators describe relaxation in the $z$-direction and dephasing in the $x$-$y$ plane of the Bloch sphere respectively. Typical relaxation and dephasing rates for flux qubits are usually of the order of 10MHz \cite{Astafiev2010}, therefore we take $\Gamma_{z}/\epsilon=\Gamma_{xy}/\epsilon=10^{-3}$. To account for the weak continuous measurement of the output field we also take $L_{b}=\sqrt{2\Gamma_{b}}\hat{b}$. From this measurement we can extract the expectation values for the position $x_{b}=\sqrt{1/2\omega_{b}}\left(b+b^{\dag}\right)$ and momentum $p_{b}=i\sqrt{\omega_{b}/2}\left(b^{\dag}-b\right)$ operators. For high quality tank circuits and transmission lines the lifetime is typically relatively long compared to the operating frequency \cite{Ilichev2003,Fink2008,Abdumalikov2008} and we therefore take $\Gamma_{b}/\omega_{b}=10^{-3}$.

	We assume that the input field has a given number of photons incident on it and is initially found in a coherent state, \ket{\alpha}, with an average of $\left|\alpha\right|^2$ photons and therefore take $f(t)=0$. We also take $h(t)=0$ and assume that the intrinsic noise in the detector system is sufficient to drive the readout field and allow detection of the incident photons. The readout field is initialised in the vacuum state and the detector qubits in the superposition state $\left(\ket{0}+\ket{1}\right)/\sqrt{2}$.

\begin{figure}
	\begin{center}
		\includegraphics[width=0.5\textwidth]{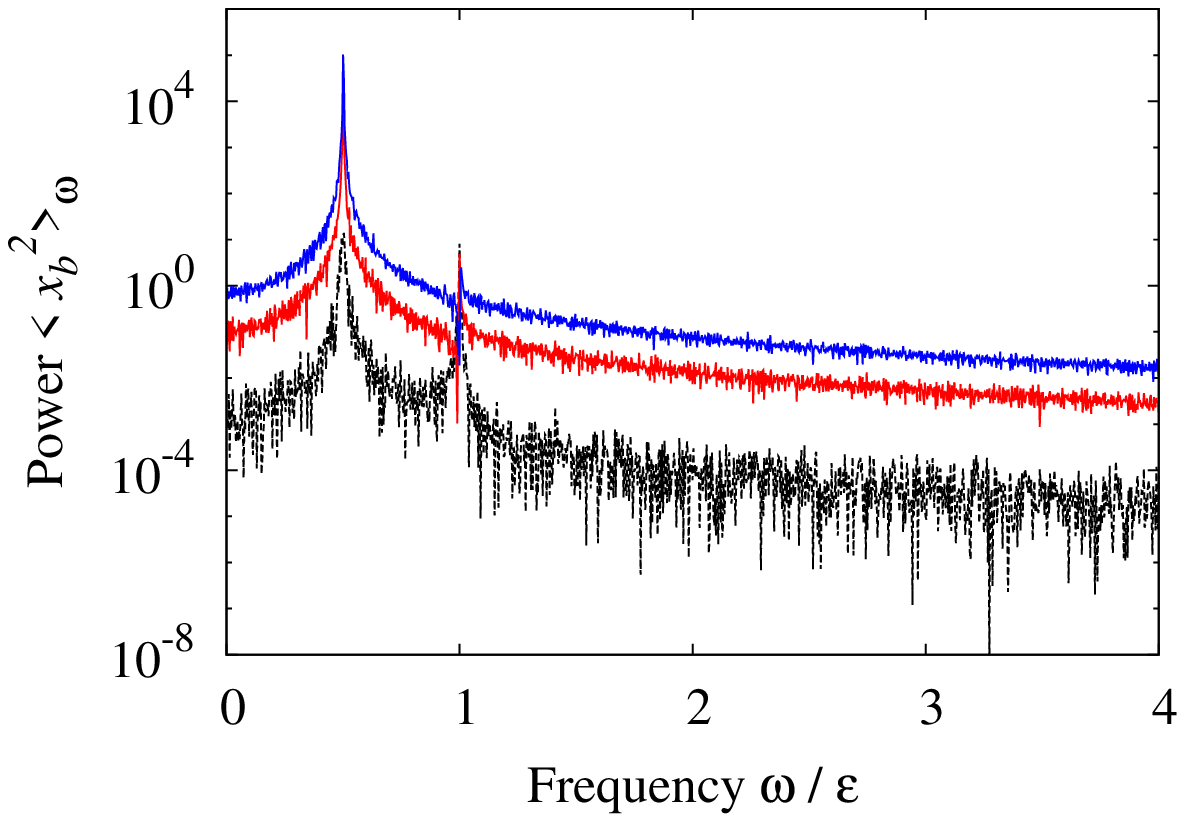}
		\includegraphics[width=0.5\textwidth]{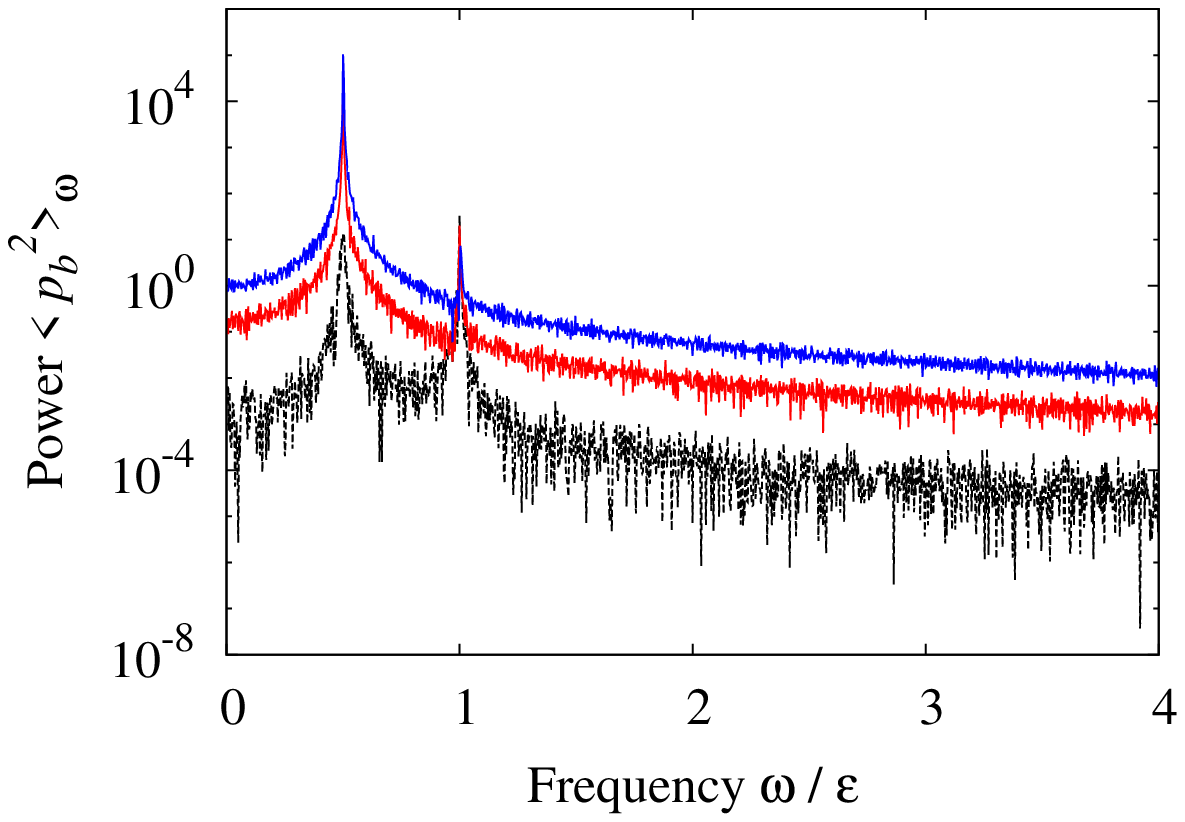}
		\includegraphics[width=0.5\textwidth]{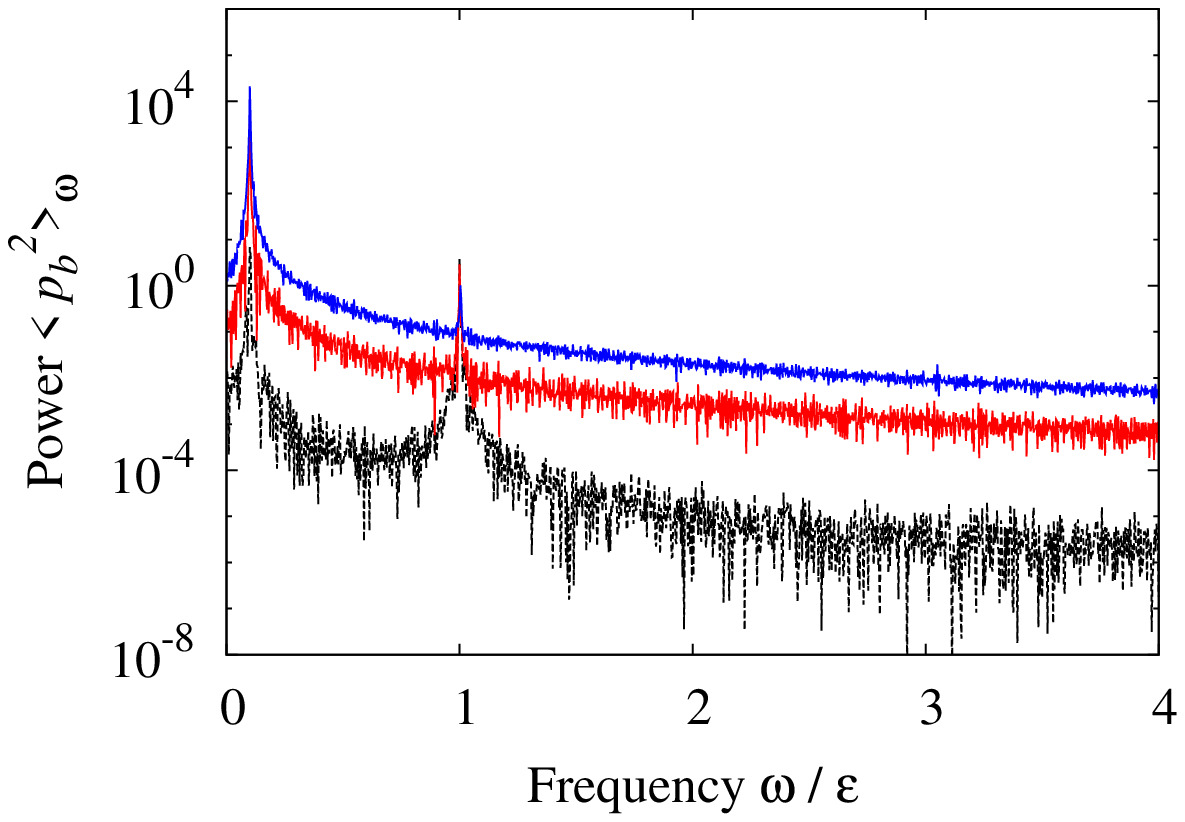}
	\end{center}
	\caption{(Color online) Examples of the readout power spectra for typical QSD realisations taken after 300 periods of $\epsilon$ where the input field has either 0 (black dashed line), 1 (red line) or 5 (blue line) photons initially incident upon it. The top and middle panes show the power spectra for the readout field's position $\left\langle x_{b}^{2}\right\rangle_{\omega}$ and momentum $\left\langle p_{b}^{2}\right\rangle_{\omega}$ quadratures for the case where $\omega_{a}/\epsilon=\omega_{b}/\epsilon=0.5$. The bottom pane shows the power spectra for the momentum quadrature $\left\langle p_{b}^{2}\right\rangle_{\omega}$ for the case of slightly mismatched field frequencies where $\omega_{a}/\epsilon=0.099$ and $\omega_{b}/\epsilon=0.1$.}
	\label{fig:powerSpectra}
\end{figure}

	Examples of power spectral densities for the readout circuit's position $\left\langle x_{b}^{2}\right\rangle_{\omega}$ and momentum $\left\langle p_{b}^{2}\right\rangle_{\omega}$ quadratures of some typical quantum trajectories are shown in \figref{fig:powerSpectra}. The key feature to note is that in each case there is a clear order of magnitude (or more) increase in the average power of the readout when a photon is incident upon the detector. We can see that this increase is still evident when there is a slight mismatch between the frequencies of the incoming photon and the readout circuit. 
	
	The difference in readout signals between a single or multiple incoming photons may not be as easily resolvable as the difference between the absence or presence of incoming photons. However, as we are interested in detecting a weak incoming signal, we can assume that the incoming photon flux is relatively low and therefore the chance of multiple photons being incident on the detector should be relatively small. It will also be possible to optimise the sampling time of the readout circuit to minimise the chance of accidentally detecting multiple photons.

%**********************************************************************************************************************************
%Numerical results by SS from here
\textit{--Scaling of the quantum metamaterial sensor array.} In order to further investigate the role of increased number of qubits and of interqubit couplings, we consider the  Hamiltonian
\begin{equation}
H = -\frac{1}{2} \sum_{j} \left[ \Delta_j \sigma^x_j + \epsilon_j(t)\sigma^z_j \right] + g \sum_j\sigma^z_j\: \sigma^z_{j+1},
\label{eq-ham}
\end{equation}
 with qubits driven by common harmonic off-resonance signal (modeling the input $V_a$ of Eq.(\ref{eq:1})) and local noise coupled through $\sigma_z$:
\begin{equation}
\epsilon_j(t)= \varepsilon \sin(\omega t) + \sqrt{2D}\xi_j(t).
\label{drive1}
\end{equation}
Here $\langle\xi(t)\rangle=0$ and $\langle \xi_j(t)\xi_l(t')=\delta_{jl}\delta(t-t')$.

\begin{figure}%
	\begin{center}
		\includegraphics[width=0.8\columnwidth]{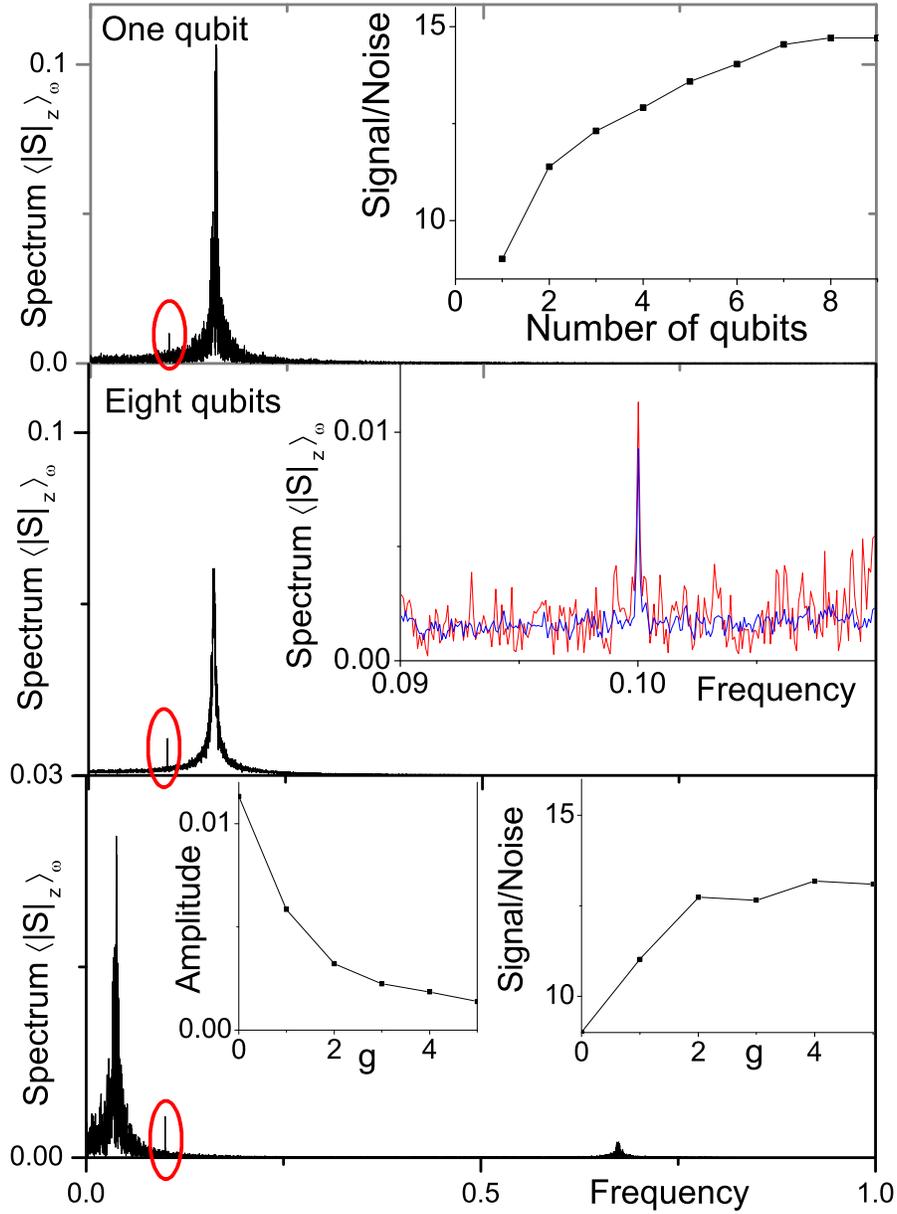}%
	\end{center}
\caption{(a) Spectral density of total detector "spin" $S_z$ in a qubit in the presence of noise and drive. The signal due to drive is a small thin peak on the left of the resonant noise response. Inset: Signal to noise ratio as the function of number of qubits. (b) Same in case of 8 qubits. Inset: A close up of the signal-induced feature. The noise is suppressed in case of 8 qubits (blue) compared to the case of a single qubit (red). (c) The spectral density of $S_z$ in case of two coupled qubits. Note that the significant shift of the resonant frequency of the system (position of the noise-induced feature). Inset: Signal response amplitude (left) and signal to noise ratio (right) as functions of the coupling strength. }%
\label{SS-fig}%
\end{figure}

In the case of $N$ uncoupled qubits, we can describe the system  by $N$ independent master equations for a single-qubit density matrix, and average the observable quantities. This data is shown in Fig. \ref{SS-fig}a,b. The spectral density of the $z$-component of the total "spin" demonstrates a small, but distinct peak due to the external drive, in addition to the large noise-driven signal. The increase of the number of qubits, predictably, increases the signal to noise ratio. The increase  is in qualitative agreement with the $\sqrt{N}$ behaviour, expected from the analytic estimate given in the Supporting materials, though numerically somewhat smaller (approximately doubling rather than tripling as $N$ increases from 1 to 9;  see Fig. \ref{SS-fig}a, inset).

% equation from our first paper with Omelianchuk
The qubit-qubit coupling also increases the signal to noise ratio. We solve the master equation for two coupled qubits,
\begin{equation}
\frac{d\hat{\rho}}{dt} = -i\left[\hat{H}(t),\hat{\rho}\right]+\hat{\Gamma}\hat{\rho},
\label{poxyj}
\end{equation}
using the generalized Bloch parametrization of the two-qubit density matrix: 
\begin{equation}
\hat{\rho}=\frac{1}{4} \sum_{a,b=0,x,y,z} \Pi_{ab} \: \sigma^1_a \otimes \sigma^2_b. \label{master}
\end{equation}
The results show that while the overall signal amplitude is suppressed by qubit-qubit coupling, the relative amplitude of the signal significantly increases (Fig. (\ref{SS-fig}c).

%**********************************************************************************************************************************
\textit{--Discussion.} Though the possibility to observe a single photon's wave-front requires the detection of a weak, remote signal against the background of local fluctuations, the standard signal-to-noise ratio $\sqrt{N}$-enhancement due to the $N$-element coherent uncoupled QMM array is unlikely to be of much practical use. Noticing that the effect of the input field is nothing but a simple one-qubit quantum gate applied to each element of the array and that introducing a simple qubit-qubit coupling scheme can improve matters, we can ponder a more sophisticated approach. By performing on a group of qubits a set of quantum manipulations, which would realize a quantum error correction routine, one can hope to improve the sensitivity of the system. We will consider this approach in a separate paper.

In conclusion, we have shown the possibility in principle to detect the wavefront of a single photon using the quantum coherent set of spatially separated qubits (a quantum metamaterial sensor array). The key feature of this approach is the combination of the nonlocal photon interaction with the collective observable of the QMM array and its QND measurement. Besides the intriguing possibility to test the limits of application of quantum mechanics, the realization of our approach would allow to greatly improve the sensitivity of radiation detectors by suppressing the effects of local noise as well as lowering the detection barrier to the minimum allowed by the uncertainty principle.

\textit{--Acknowledgements.} The authors are grateful to F.V. Kusmartsev for stimulating discussions. AMZ, RDW, ME and SS acknowledge the support of the John Templeton Foundation. DRG and VKD were partially supported by the project "Development of ultrahigh sensitive receiving systems of THz wavelength
range for radio astronomy and space missions" in NSTU n.a. R.E. Alekseev. EI acknowledges funding from  the European Community's Seventh Framework Programme (FP7/2007-2013) under Grant No. 270843 (iQIT).
 
\begin{figure}%
\includegraphics[width=0.5\columnwidth]{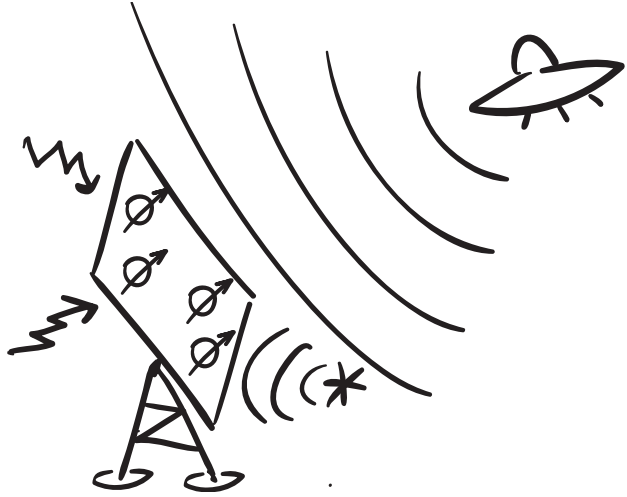}%
\caption{The wave front of a single photon originating from a remote source being detected by spatially separated qubits (a quantum metamaterial-based sensor array) against uncorrelated background noise (nearby radiation sources and local fluctuations).}%
\label{fig:saucer}%
\end{figure}

%**********************************************************************************************************************************
\textit{--Supporting materials.} For a simple illustration of how a single photon can be simultaneously detected at several points in space (Fig. \ref{fig:saucer}), consider the case when there is one photon in the input circuit,  two identical and identically coupled to it noiseless qubits  are initially in their ground state, and the readout circuit is switched off. In this case the system undergoes vacuum Rabi oscillations,  and its wave function is \cite{Smirnov2002}
\begin{eqnarray}
|\Psi(t)\rangle = \cos(\sqrt{2}g^at) |1\rangle \otimes |\downarrow_1\rangle \otimes |\downarrow_2\rangle + \nonumber \\
i \sin(\sqrt{2}g^at) |0\rangle \otimes \frac{|\downarrow_1\rangle \otimes |\uparrow_2\rangle + |\uparrow_1\rangle \otimes |\downarrow_2\rangle}{\sqrt{2}}.
\label{eq:2}
\end{eqnarray}
At the moments when the first term vanishes, $t_n = (\pi/2 + \pi n)/\sqrt{2}g^a,$  the qubits are in the maximally entangled Bell state, and the QND measurement of their summary "spin" in z-direction realizes the observation of a single photon's presence (a Fock state $|1\rangle$ of the circuit A)  at two spatially separated points (locations of the qubits 1 and 2).

A literal realization of such a  scheme for observing a single photon's wavefront in multiple points is theoretically possible, but hardly advisable: The resonant transfer of the incoming photon into the qubit array and back is vulnerable to absorption in one of the qubits. A better opportunity is presented by the dispersive regime, when the mismatch between the qubits' and incoming photon's resonant frequencies, $\delta\Omega_j = |\omega_a - \sqrt{\Delta_j^2 + \epsilon_j^2}| \gg g^a_j$, allows to use the Schrieffer-Wolff transformation. If $\Delta = 0$, the interaction term $V_a$ in (\ref{eq:1}) is reduced to \cite{Blais2004,Zagoskin2011}
\begin{equation}
\tilde{V}_a = \left( \sum_j \frac{(g^a_j)^2}{\delta\Omega_j} \sigma^z_j \right) a^{\dag}a.
\label{eq:after-SW}
\end{equation}
Now the effect of the input field on the detector qubits is the additional phase gain proportional to the number of incoming photons, which can be read out using a QND technique.  

Note that in addition to  (\ref{eq:after-SW}) we will also obtain the effective coupling between qubits through the vacuum mode of the oscillator (see, e.g., \cite{Makhlin2001}). In case of identical qubits and coupling parameters it is 
\begin{equation}
\tilde{H}_{\rm eff} = \frac{(g^a)^2}{2\delta\Omega} \sum_{jk} \sigma^x_j\sigma^x_k.
\label{eq:almost}
\end{equation}
If the number of qubits in the matrix is large enough, this term can be approximated by a "mean field" producing an effective "tunneling" for each individual qubit, $\Delta_{\rm eff}(t)\sigma^x_j = \langle \sum_{k} \sigma^x_k\rangle \sigma^x_j.$

In the following, it will be convenient to account for the ambient noise sources through the term 
\begin{equation}
H_{\rm noise} = \sum_j \left( \xi_j(t) \sigma^x_j + \eta_j(t) \sigma^z_j \right)
\label{eq:Hnoise}
\end{equation}
in the Hamiltonian. In agreement with our assumptions, these fluctuations in different qubits are uncorrelated: $\langle\xi_j(t)\xi_k(t')\rangle \propto \delta_{jk}; \langle\xi_j(t)\delta\eta_k(t')\rangle = 0$).    

Let us excite the input circuit with a resonant field, $f(t) = f_e(t)\exp[-i\omega_at] + c.c.$, with slow real envelope function $f_e(t)$. Neglecting for the moment the rest of the system, due to the weakness of the effective coupling $g^2/\delta\Omega$ in (\ref{eq:after-SW}), we can write for the wave function of the input circuit
\begin{equation}
i \frac{d}{dt}|\psi_a(t)\rangle \approx f_e(t) (a + a^{\dag}) |\psi_a(t)\rangle,
\label{eq:a1}
\end{equation}
and
\begin{equation}
|\psi_a(t)\rangle \approx e^{-i\left[\int_0^t dt' f_e(t')\right](a + a^{\dag})} |\psi_a(0)\rangle \equiv D(\alpha)|\psi_a(0)\rangle.
\label{eq:a2}
\end{equation}
Here $D(\alpha)$ with 
\begin{equation}
\alpha(t) = -i\left[\int_0^t dt' f_e(t')\right],
\label{eq:D2}
\end{equation} 
is the displacement operator 
\begin{equation}
D(\alpha) = e^{\alpha a^{\dagger} - \alpha^* a}.
\label{eq:D1}
\end{equation}
Acting on a vacuum state, it produces a coherent state, $ D(\alpha)|0\rangle = |\alpha\rangle. $ Therefore, assuming that the input circuit was initially in the vacuum state, the average
\begin{equation}
\langle a^{\dagger}a \rangle_t \approx \langle \psi_a(t)| a^{\dagger}a |\psi_a(t)\rangle \approx \langle\alpha(t)|a^{\dagger}a|\alpha(t)\rangle = |\alpha(t)|^2 = \left[\int_0^t dt' f_e(t')\right]^2.
\label{eq:Na}
\end{equation}
Therefore the action of the incoming field on the qubits in the dispersive regime can be approximated by replacing the terms $H_a$ and $V_a$ in the Hamiltonian (\ref{eq:1}) with
\begin{equation}
h(t) = \left( \sum_j \frac{(g^a_j)^2}{\delta\Omega_j} \sigma^z_j \right) |\alpha(t)|^2 \equiv \left( \sum_j \gamma_j \sigma^z_j \right) |\alpha(t)|^2.
\label{eq:after-coherent}
\end{equation}

In the Heisenberg representation the "spin" of the $j$th qubit, 
\begin{equation}
\vec{s}_j = s^x_j \sigma^x_j + s^y_j \sigma^y_j + s^z_j \sigma^z_j, 
\label{eq:spin}
\end{equation} 
satisfies the Bloch equations, which in case of zero bias and only $z$-noise $(\epsilon_j = 0; \xi_j(t) = 0)$, and neglecting for the moment the interaction with the readout circuit, take the form
\begin{eqnarray}
\frac{d}{dt}s^x_j(t) = 2 [\gamma_j |\alpha(t)|^2 + \eta_j(t)] s^y_j(t); \nonumber \\
\frac{d}{dt}s^y_j(t) = - 2 [\gamma_j |\alpha(t)|^2 + \eta_j(t)] s^x_j(t) - \Delta_{\rm eff} s^z_j(t);  \\
\frac{d}{dt}s^z_j(t) =  \Delta_{\rm eff} s^y_j(t), \nonumber
\label{eq:Bloch}
\end{eqnarray}
or, introducing $s^{\pm}_j = s^{x}_j \pm i s^{y}_j$,
\begin{eqnarray}
\frac{d}{dt}s^{\pm}_j(t) = \mp\left\{ 2i [\gamma_j |\alpha(t)|^2 + \eta_j(t)] s^{\pm}_j(t) + i\Delta_{\rm eff} s^z_j(t)\right\}; \nonumber \\
\frac{d}{dt}s^z_j(t) =  \frac{\Delta_{\rm eff}}{2i} \left[ s^+_j(t) - s^-_j(t) \right]. 
\label{eq:Bloch2}
\end{eqnarray}

Let us initialize the qubit in an eigenstate of $\sigma^x_j$ (i.e., in an eigenstate of unperturbed qubit Hamiltonian, since $\epsilon_j = 0$). Then $s^z_j(0) = 0, s^{\pm}_j(0) = s^x_j(0)$ (i.e., 1 or -1), and, assuming the slowness of $\Delta_{\rm eff}(t)$, the equations (\ref{eq:Bloch2}) can be solved perturbatively:
\begin{eqnarray}
s^{\pm (0)}_j(t) = \exp\left[\mp 2i \int_0^t [\gamma_j |\alpha(t')|^2 + \eta_j(t')] dt'\right]  s^x_j(0); \nonumber \\
s^{z (1)}_j(t) = - \Delta_{\rm eff} s^x_j(0) \int_0^t \sin  \left\{ 2 \int_0^{t'} [\gamma_j |\alpha(t'')|^2 + \eta_j(t'')] dt''\right\}  dt'
\approx \\
- 2 \Delta_{\rm eff} s^x_j(0) \int_0^t    \int_0^{t'} [\gamma_j |\alpha(t'')|^2 + \eta_j(t'')] dt' dt''. \nonumber
\label{eq:approx1}
\end{eqnarray}
Assuming identical qubits identically coupled to the input circuit, we finally obtain for the collective variable (z-component of the total qubit "spin" of the QMM array)
\begin{equation}
S^z(t) \equiv \sum_{j=1}^N s^z_j(t) \approx -2  \gamma \Delta_{\rm eff} s^x(0) N \left[  \int_0^t\int_0^{t'} |\alpha(t'')|^2  dt' dt''  + \int_0^t\int_0^{t'}  \frac{1}{N}\sum_{j=1}^N \eta_j(t'')  dt' dt''\right].  
\label{eq:approx2}
\end{equation} 
The second term in the brackets is the result of local fluctuations affecting separate qubits and is therefore, in the standard way, $\sim \sqrt{N}$ times suppressed compared to the first term (due to the regular evolution produced by the spatially coherent input photon field). The variable $S^z$ can be read out by the output LC circuit, e.g., by monitoring the equilibrium current/voltage noise in it \cite{Ilichev2003}. The signal will be proportional to the spectral density $\langle \left( S^z \right)^2 \rangle_{\omega}$, i.e. to the Fourier transform of the correlation function $\langle S^z(t+\tau)S^z(t)\rangle$. Due to the quantum regression theorem \cite{Gardiner}, the relevant correlators satisfy the same equations (\ref{eq:Bloch2}) as the operator components themselves, and the "regular" and "noisy" terms originating from (\ref{eq:approx2}) will indeed be $O(N^2)$ and $O(N)$ respectively.

\bibliographystyle{unsrt}
\bibliography{paramRefs616}
%\bibliography{detectorRefs-v2}
\end{document}